\documentclass[11pt]{article}

\usepackage[margin=1in]{geometry}
\usepackage{times}
\usepackage[T1]{fontenc}
\usepackage[utf8]{inputenc}
\usepackage{graphicx}
\usepackage{booktabs}
\usepackage{array}
\usepackage{amsmath}
\usepackage{authblk}
\usepackage[colorlinks=true,linkcolor=blue,citecolor=blue,urlcolor=blue,breaklinks=true]{hyperref}
\usepackage{xurl}
\usepackage[round]{natbib}
\usepackage{setspace}
\usepackage{caption}
\usepackage{titlesec}

\setcitestyle{authoryear,open={(},close={)}}
\title{\textbf{From Caveman to Expert Analyst: Energy Consumption of Variable LLM Tasks}}

\author[1]{Diego Manya}
\author[2]{Ethan I. Thorpe}
\author[3]{Ji Zhang}
\author[4]{Myranda Shirk}
\author[3]{Jiamian He}
\author[1,5]{Angel Hsu\thanks{Corresponding author: \href{mailto:angel.hsu@unc.edu}{angel.hsu@unc.edu}. Postal address: 123 W. Franklin Street, Suite 330b, Chapel Hill, NC 27516. Phone: 864.354.3499.}}
\author[6]{Michael P. Vandenbergh}

\affil[1]{Data-Driven EnviroLab, Institute for Environment, UNC Chapel Hill}
\affil[2]{Climate Governance Lab, Vanderbilt Law School, Nashville, Tennessee, USA}
\affil[3]{Arboretica, Rotterdam, The Netherlands}
\affil[4]{Vanderbilt University, Nashville, Tennessee, USA}
\affil[5]{Department of Public Policy, UNC Chapel Hill}
\affil[6]{Vanderbilt Law School, Nashville, Tennessee, USA}

\date{}

\begin{document}

\maketitle

\begin{abstract}
\noindent The energy demand growth and environmental impacts of artificial intelligence (AI) have generated substantial interest in supplying sufficient low-cost electricity for AI-driven data center development. Research on the ability of demand-side management to address these challenges has been more limited. Shifting the amount or timing of demand from retail, corporate, and other organizational behaviors is a plausible option but only if changes in demand-related behavior have important effects on the environmental and electricity effects of AI. This article tests four retail (i.e., consumer) user behaviors with high behavioral plasticity to assess their technical abatement potential. The research concludes that non-reasoning models provide sufficient quality while consuming close to one-twentieth of energy compared to reasoning models, saving an amount equal to the annual electricity requirement of at least 141,000 US households under daily usage assumptions. Simple prompt modifications can yield additional reductions in energy consumption by up to 65\% using non-reasoning models. Specifically, the practice that maintains the highest degree of similarity with the baseline reduces electricity demand in the range of 4 to 35\%, an amount equal to the annual electricity requirement of up to 7,200 US households. Although AI advancements make precise estimates of environmental and electricity impacts difficult to assess, the results confirm that certain minimally intrusive best practices aimed at the majority of users can reduce the energy and environmental burdens imposed by AI.

\vspace{0.5em}
\end{abstract}

% ============================================================
\section{Introduction}
% ============================================================

Generative AI (GenAI) development and usage, particularly the usage of large language model (LLM) chatbots, has led to substantial projected energy demand growth and environmental impacts. As of June 2026, ChatGPT alone reported surpassing a billion users \citep{Chin2026} and Google reported increasing its AI load seven-fold relative to the prior year, reaching more than three quadrillion tokens per month \citep{Pichai2026}. Although the International Energy Agency (IEA) estimates that data centers only comprised 1.5\% of global electricity usage in 2024, global usage is expected to double by 2030 \citep{Cozzi2025}. Inference (e.g., individual usage or AI prompting) is projected to grow by 30\% annually as AI adoption accelerates, and to account for half of the projected growth in electricity consumption between 2024 and 2030 \citep{Tanner2026}. Unlike traditional LLMs, Agentic AI involves repeated, iterative inferences, entailing much greater energy consumption and associated carbon emissions, depending on the design and usage of the GenAI application \citep{Li2024,Bao2026,Oviedo2026}. The rapid proliferation of AI, therefore, has upended decades of consistent electricity demand, challenging reliability, affordability, and environmental goals.

Research and policy have focused on supplying low-cost, and in some cases low environmental impact, electricity for data centers \citep{FERC2026,He2024}. GenAI is used daily by almost everyone with an internet connection (e.g., Google AI overview) \citep{Maese2025}, although only 55\% of people around the world and 44\% of Americans know which services use AI \citep{Ipsos2026}. Its growth has been largely attributable to the rise of LLMs like ChatGPT, Gemini, and Claude that have generated widespread public awareness of and demand for AI tools. At the same time, research suggests that retail user knowledge about the energy and environmental footprint of AI is very limited. For instance, in a recent study roughly a third of United States survey respondents were unaware or only slightly aware that AI uses electricity, and almost half were unaware of environmental impacts. Opportunities for demand-side management have received less attention. Although demand-side opportunities arising from the ability to use contracts to shift the timing of demand among AI firms, data centers, and utilities have demonstrated promise \citep{Norris2025}, less research has been conducted on the potential for changes in user behavior to occur and have substantial effects on the timing and amount of electricity use. If shifting the amount or timing of demand from ``retail'' (i.e., casual individual user), corporate, and other organizational behaviors can be achieved, there is a potential for demand-side behavior to have important effects on AI electricity demand \citep{Allen2026}.

Here, we examine how modifications in prompting behavior (i.e., text-based instructions or queries given to an LLM to generate a desired response) \citep{Podder2026} shape AI electricity demand. In doing so, we analyze various prompting approaches individuals can employ to reduce energy demand from their GenAI use, and that organizations can include in policies or recommendations for their employees and others using AI to reduce costs. While prior studies \citep{Podder2026,Adamska2026,Rubei2025} have examined the relationship between prompt engineering and energy consumption or carbon emissions, they primarily focus on open-source models (e.g., Llama) or specialized use cases, such as technical writing, content creation, coding or data analysis, rather than a ``retail'' user who is utilizing a commercially available generic LLM such as ChatGPT or Claude. Furthermore, this study expands upon other studies by analyzing results by task complexity, highlighting how the effects of prompt modification differ depending on the use case. While additional environmental impacts associated with GenAI, such as water consumption \citep{Li2024,Han2026} and carbon emissions \citep{Luccioni2024} can potentially be reduced by demand-side management, our research only assessed electricity usage because water consumption and emissions are highly variable and dependent on conditions outside of user control (e.g., temperature in the data center location or electricity mix of the grid serving a data center).

To have a substantial impact on energy demand, large numbers of individual GenAI users will need to make these changes, a shift that may require new informational initiatives by some combination of GenAI firms, employers, schools, non-profit groups, and others. Assessments of behavior change opportunities often include evaluation of the technical potential (TP) of potential behaviors (the energy use reduction from the behavior), as well as the behavioral plasticity (BP) of these behaviors (the ease of behavior change) \citep{Dietz2009,Swim2025,Nielsen2026}. This study differentiates itself from previous work by researching universally reusable prompt techniques (i.e., ones that can be copied and pasted into any prompt) that are more accessible for many users. The rapid change in GenAI suggests that best practices for lowering energy use may change over time. Although the specific approaches for reducing energy demand evaluated in this research may shift, the goal of the research is to evaluate whether individual behaviors provide a basis for interventions and identify behavior changes that will result in substantial shifts in the timing and amount of electricity demand.

This research explores whether those behaviors exist and what their impact might be in the context of individual commercial LLM use. We focus on testing simple, user-accessible practices that require no technical expertise, such as selecting between reasoning and non-reasoning models, prompting with various ``personas'' (e.g., an ``energy-efficient'' persona), requesting minimal responses, and using highly efficient or truncated prompting. To evaluate these practices, we deployed stratified prompts by task complexity using Bloom's Taxonomy and selected prompts from LLM benchmarking datasets on the latest commercial models from major providers \citep{Brynjolfsson2026}.

% ============================================================
\section{Data and Methods}
% ============================================================

Evaluating the energy implications of LLM use depends not only on model architecture, hardware, and data-center operations \citep{Samsi2023,Fernandez2025}, but also on how users interact with commercial LLM-based chatbot systems. Whether ordinary or retail users can reduce query-level energy demand through actions such as model selection or prompting practices is the research question in this study. When an LLM generates a response to a prompt, it is generated through several stages. The model first processes a user's input prompt and then generates the response incrementally, adding one token (e.g., roughly a word or part of a word) at a time \citep{Podder2026}. The resulting energy required therefore depends partly on the length of the input prompt, but also on the length and complexity of the response elicited. Prior research suggests that output length is especially important, since studies have shown a strong correlation between the output length and energy consumption \citep{Adamska2026,Podder2026}.

The content and structure of prompts have also been found to affect energy consumption. \citet{Adamska2026}, for instance, find variable LLM energy consumption based on certain key words that often correlate with the complexity of the task, particularly when the prompting language encourages more concise outputs, such as ``classify'' rather than ``explain.'' Similarly, \citet{Rubei2025} show that prompt-engineering strategies such as ``zero-shot'' prompting versus one-shot or few-shot approaches that add example responses to the input context affect energy consumption. Together, these studies suggest that prompting practices may influence energy use by changing both the amount of input the model must process and the amount of generated input, although they primarily evaluated open-source only LLMs like Llama and Mistral, since these models can be deployed locally and combined with software that allows for more direct measurement of energy consumption. To date, few studies have evaluated energy consumption of commercial LLM chatbots, where information regarding energy consumption of specific models and AI tasks is more limited.

Building on these prior studies, our study examines whether prompt- and model-level choices produce measurable differences in energy consumption under realistic commercial LLM usage. We focus on ten models provided by five of the most popular AI companies (OpenAI, Anthropic, xAI, Google, and DeepSeek; Table~\ref{tab:models}) and practices that ordinary or ``retail'' users can implement directly, such as selecting between reasoning and non-reasoning models or modifying prompting language. Because the effects of these practices may vary by the cognitive demands of the task, we also stratify prompts using Bloom's taxonomy \citep{Bloom1956}, allowing us to assess whether energy-saving potential differs across knowledge, comprehension, application, analysis, synthesis and evaluation tasks.

\begin{table}[htbp]
\centering
\caption{LLM models and parameters used for evaluation. \textit{Note}: Parameter values are estimated from \citet{Li2026} due to undisclosed information from providers. Gemini 3.1 Pro and 3.1 Flash are lower-bound estimates using Gemini 2.5 Pro and 2.5 Flash, respectively.}
\label{tab:models}
\small
\begin{tabular}{@{}p{1.6cm}p{2.6cm}p{2.1cm}p{1.5cm}p{5.2cm}@{}}
\toprule
\textbf{Provider} & \textbf{Model} & \textbf{Type} & \textbf{Parameters} & \textbf{Inference configuration} \\
\midrule
OpenAI & gpt-5.5-pro & Reasoning & 9.7T & temp=0, reasoning\_effort=``high'' \\
OpenAI & gpt-5.4-mini & Non-reasoning & 410B & temp=0, reasoning\_effort=``low'' \\
Google & gemini-3.1-pro-preview & Reasoning & 1.2T & temp=0, thinking\_level=``high'' \\
Google & gemini-3.1-flash-lite & Non-reasoning & 70B & temp=0, thinking\_level=``minimal'' \\
Anthropic & claude-opus-4-7 & Reasoning & 4.0T & temp=1, thinking=``adaptive'' \\
Anthropic & claude-haiku-4.5 & Non-reasoning & 65B & temp=1, thinking=``enabled'',\\ & & & & budget\_tokens=4096 \\
DeepSeek & deepseek-v4-pro & Reasoning & 16T & temp=0, thinking=``enabled'',\\ & & & & reasoning\_effort=``high'' \\
DeepSeek & deepseek-v4-flash & Non-reasoning & 284B & temp=0, thinking=``disabled'' \\
xAI & grok-4.3 & Reasoning & 3.2T & temp=0, reasoning\_effort=``high'' \\
xAI & grok-4.3 & Non-reasoning & 3.2T & temp=0, reasoning\_effort=``low'' \\
\bottomrule
\end{tabular}
\end{table}

\subsection{Testing Methodology and Data Analysis}

We identified 50 prompts in each of the six Bloom's taxonomy categories (Knowledge, Comprehension, Application, Analysis, Synthesis, and Evaluation) for a total of 300 queries from existing LLM benchmarking datasets such as ChatBot Arena, Natural Questions, and MMLU \citep{Zheng2023,GoogleResearch2021,Hendrycks2021}. Selected queries were based on real user input with LLMs. In addition to the overall effect of the behaviours on energy, we use Bloom's taxonomy to explore whether the energy reduction effects of the evaluated practices vary across queries of increasing complexity. In order to evaluate semantic similarity between the models, we performed a cosine similarity between model pairs. We use the all-MiniLM-L6-v2 sentence transformer model \citep{Reimers2019} to map the answers to vector space and the scikit-learn Python library for the cosine similarity function \citep{Pedregosa2011}.

We then modified the prompts with three simple prompting techniques designed to reduce energy consumption without negatively affecting response quality. The baseline scenario---non-reasoning models without prompt modification---mirrors the default configuration in most commercially available LLMs. The four evaluated practices explore two different types of modification to the baseline scenario. The first practice assesses energy consumption using the latest reasoning or high-effort model while keeping the prompts unchanged. The next three practices alter the prompts while keeping the model the same as the baseline scenario (i.e., a base prompt using the most recent non-reasoning or low-effort models). These practices are not meant to be exhaustive, but rather a selection of potential modifications that can be implemented in commercial LLMs by retail users regardless of technical capacity, as opposed to more technical modifications implemented in open-source models or in LLM applications \citep{Kuran2026,Caravaca2025}. The details of the baseline scenario and the evaluated practices are described in Table~\ref{tab:practices}.

\begin{table}[htbp]
\centering
\caption{Baseline and practices selected for analysis.}
\label{tab:practices}
\small
\begin{tabular}{@{}p{2.8cm}p{10.5cm}@{}}
\toprule
\textbf{Practices} & \textbf{Description} \\
\midrule
Baseline & Base prompt using most recent non-reasoning or low-effort models. \\[4pt]
High-reasoning & Base prompt using most recent reasoning and high-effort models. \\[4pt]
Energy-efficient persona & Base prompt modified to include the following instruction: ``You are an energy efficient LLM designed to minimize energy consumption from your use without reducing response quality.'' Used with non-reasoning or low-effort models. \\[4pt]
Minimal answer & Base prompt modified to include the following instruction: ``Respond using minimal tokens to answer my question completely and accurately. Do not include filler.'' Used with non-reasoning or low-effort models. \\[4pt]
Caveman & Base prompt modified to include the following instruction: ``Respond terse like smart caveman --- drop articles, filler, pleasantries. Fragments OK. Technical terms exact. Code unchanged. Pattern: [thing] [action] [reason]. [next step].'' \citep{Brussee2026}. Used with non-reasoning or low-effort models. \\
\bottomrule
\end{tabular}
\end{table}

Unlike previous studies that estimate energy efficiency for open-source LLMs run on researchers' hardware (e.g., \citealp{Luccioni2022}), we focused on estimating the energy consumption of the commercially available LLMs as they are commonly used by retail users. The details of those models and the simplified equations are detailed in Table~\ref{tab:energy}. Due to the limited transparency from commercial LLM providers to disclose real-time server-side energy consumption \citep{Vandenbergh2026}, our energy estimations build upon \citet{Jegham2025}'s methodology modified by \citet{Bao2026} for single query energy estimation. Due to the use of client-side time to last token (TTLT) for each query, we executed each request from the US during nighttime (local server time) in order to reduce the effect of location and queuing in total response time. For each query, we recorded token use and the response time which is used to estimate energy consumption.

Energy estimation is calculated for each model based on the specific parameters identified by \citet{Jegham2025} with updated data for newer models and assumptions regarding batching considering 32 concurrent batch processing for non-reasoning models and 16 for reasoning models \citep{Anyscale2023}. The energy values are calculated using total response time multiplied by a factor that represents the theoretical GPU and non-GPU energy consumption for each model based on their power specifications and Power Usage Effectiveness (PUE).

\begin{table}[htbp]
\centering
\caption{Simplified energy consumption calculation for each model.}
\label{tab:energy}
\small
\begin{tabular}{@{}p{1.6cm}p{3.5cm}p{2.1cm}p{7cm}@{}}
\toprule
\textbf{Provider} & \textbf{Model} & \textbf{Type} & \textbf{Simplified energy equation} \\
\midrule
OpenAI & gpt-5.5-pro & Reasoning & $E(\text{Wh}) = 4872.0 \times (\text{inference\_time}/3600)$ \\
OpenAI & gpt-5.4-mini & Non-reasoning & $E(\text{Wh}) = 410.4 \times (\text{inference\_time}/3600)$ \\
Google & gemini-3.1-pro-preview & Reasoning & $E(\text{Wh}) = 328.4 \times (\text{inference\_time}/3600)$ \\
Google & gemini-3.1-flash-lite & Non-reasoning & $E(\text{Wh}) = 74.0 \times (\text{inference\_time}/3600)$ \\
Anthropic & claude-opus-4-7 & Reasoning & $E(\text{Wh}) = 1627.3 \times (\text{inference\_time}/3600)$ \\
Anthropic & claude-haiku-4.5 & Non-reasoning & $E(\text{Wh}) = 203.4 \times (\text{inference\_time}/3600)$ \\
DeepSeek & deepseek-v4-pro & Reasoning & $E(\text{Wh}) = 1485.8 \times (\text{inference\_time}/3600)$ \\
DeepSeek & deepseek-v4-flash & Non-reasoning & $E(\text{Wh}) = 247.7 \times (\text{inference\_time}/3600)$ \\
xAI & grok-4.3 & Reasoning & $E(\text{Wh}) = 2299.4 \times (\text{inference\_time}/3600)$ \\
xAI & grok-4.3 & Non-reasoning & $E(\text{Wh}) = 1149.7 \times (\text{inference\_time}/3600)$ \\
\bottomrule
\end{tabular}
\end{table}

% ============================================================
\section{Results}
% ============================================================

\subsection{Reasoning vs.\ non-reasoning model comparison}

\begin{figure}[htbp]
\centering
\includegraphics[width=0.85\textwidth]{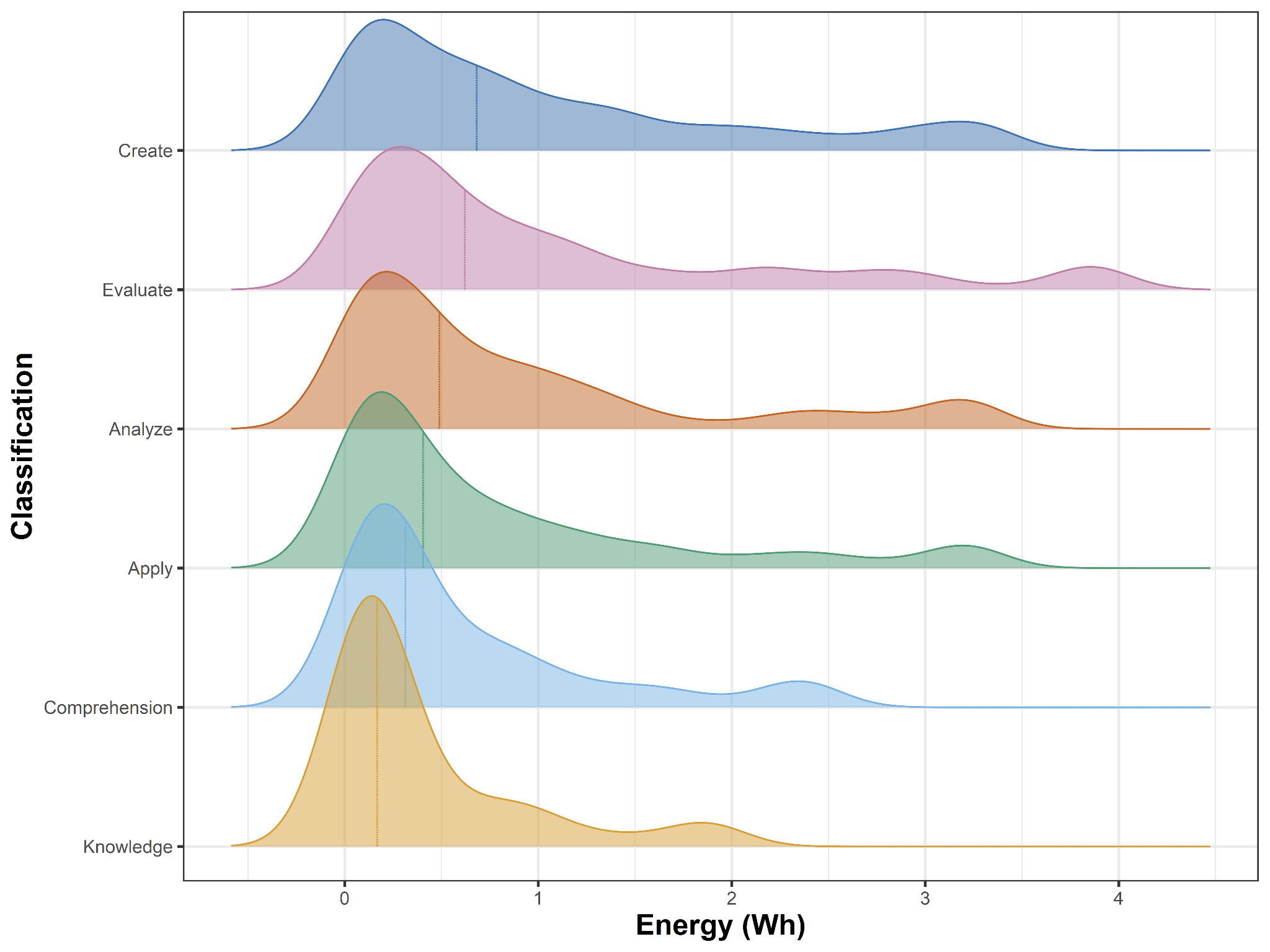}
\caption{Energy usage (Wh) distribution by classification type according to Bloom's taxonomy.}
\label{fig:1}
\end{figure}

The results of our baseline estimations (non-reasoning models) in Figure~\ref{fig:1} show a gradient in the required energy following the complexity of the prompts with the lowest median estimates (0.16 Wh) for knowledge type prompts and highest median estimates (0.68 Wh) for create type of prompts. Further statistical tests on the difference between these distributions (Wilcoxon rank tests) show that lower cognitive tasks (Knowledge, Comprehension and Apply) are statistically different in energy consumption than higher cognitive tasks (Evaluate and Create), while mid-level tasks (Apply and Analyze) show less differentiation between them in their energy usage. Our estimates by model also show this gradient in energy consumption across all the five models used for the baseline estimations (Figure~\ref{fig:s1}). Overall, this result highlights both the usefulness of Bloom's taxonomy for classifying general use prompts in LLM's energy usage as well as the existence of a statistically significant difference in energy consumption as we move from low to high cognitive complexity questions.

\begin{figure}[htbp]
\centering
\includegraphics[width=0.85\textwidth]{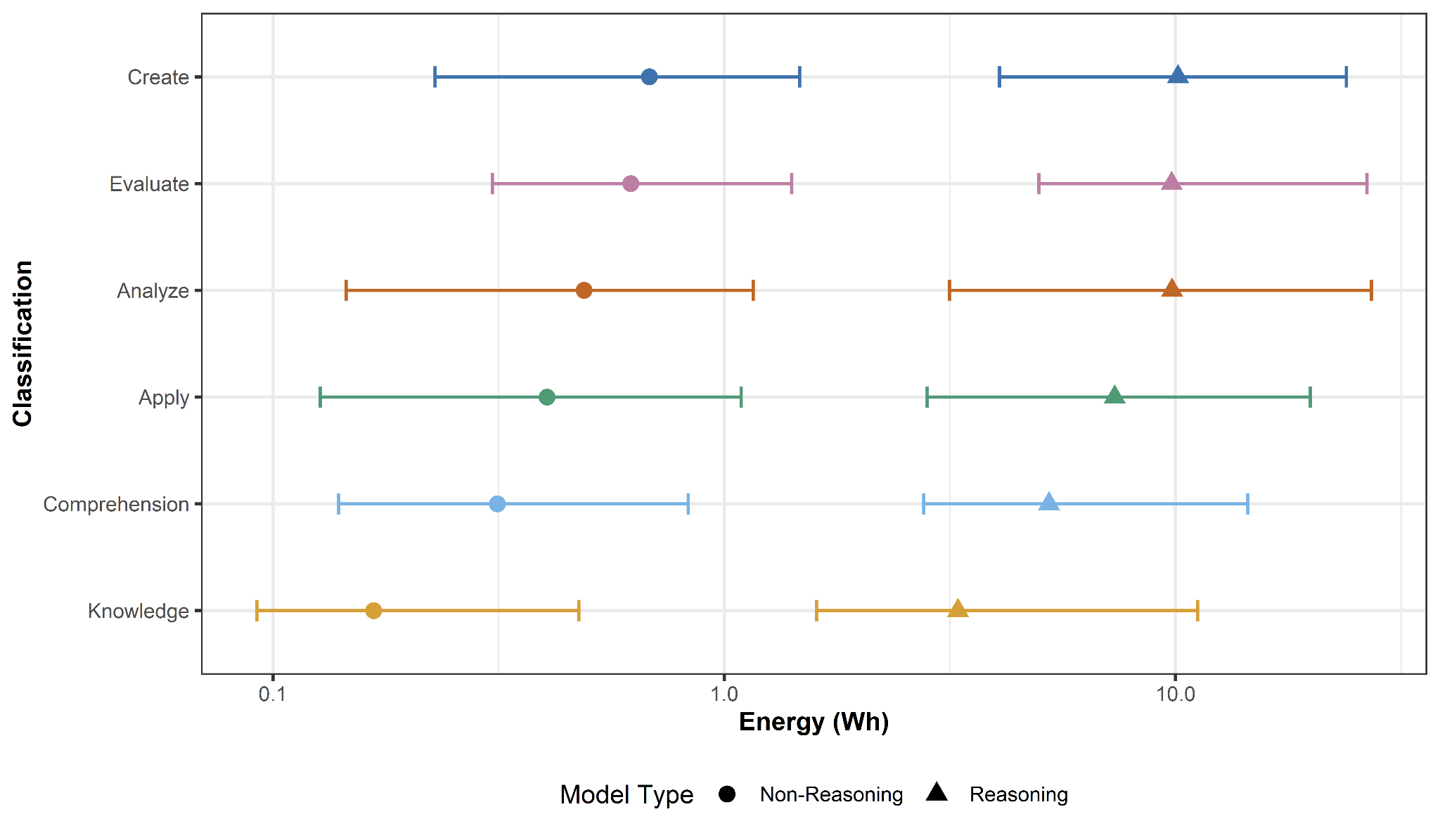}
\caption{Comparison of energy use for non-reasoning and reasoning models across questions classified according to Bloom's taxonomy.}
\label{fig:2}
\end{figure}

When comparing the results of our baseline to the first evaluated practice, using a high-reasoning model, we observe that the baseline consumes substantially less energy across all task categories (Figure~\ref{fig:2}). Reasoning models exhibit a similar relationship between energy consumption and cognitive task complexity, but at a consistently higher level of energy use. The relative difference is largest for the lowest-complexity tasks, where reasoning models consume approximately 20 times more energy than the baseline, and narrows for higher-complexity tasks, where they consume approximately 15 times more energy. These patterns are observed when comparing each non-reasoning model with its reasoning counterpart from the same provider (Figure~\ref{fig:s2}).

\begin{figure}[htbp]
\centering
\includegraphics[width=0.85\textwidth]{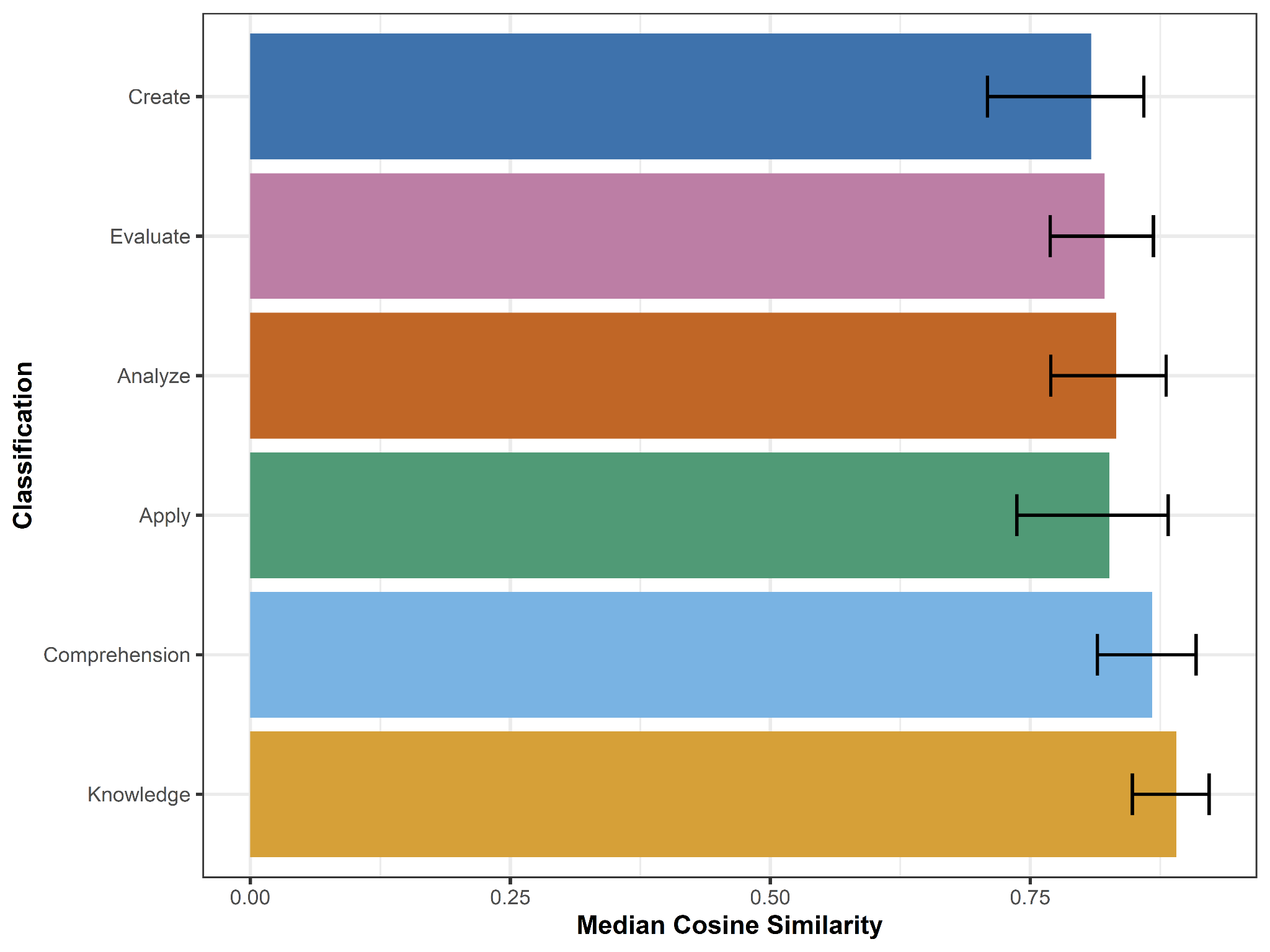}
\caption{Comparison of semantic similarity for non-reasoning models against reasoning model output.}
\label{fig:3}
\end{figure}

To understand whether the increase in energy consumption was associated with differences in response quality, we analyzed the cosine similarity between each pair of answers using the non-reasoning and reasoning models of the same developer (Figure~\ref{fig:3}). Our results find that the semantic similarity between the answers for all question types is very high (score $>$ 0.75), suggesting that while the energy cost is at least ten times higher for reasoning models, the difference in semantic content between the non-reasoning and reasoning models is very small (10\% in Knowledge increasing to 20\% in Creation tasks). Therefore, for most use cases except those that are exceptionally complex or demand exceptional precision, switching from a non-reasoning to a reasoning model will not see a significantly better response but will increase their energy footprint by at least 19 times.

\subsection{Comparing prompting practices}

\begin{figure}[htbp]
\centering
\includegraphics[width=0.85\textwidth]{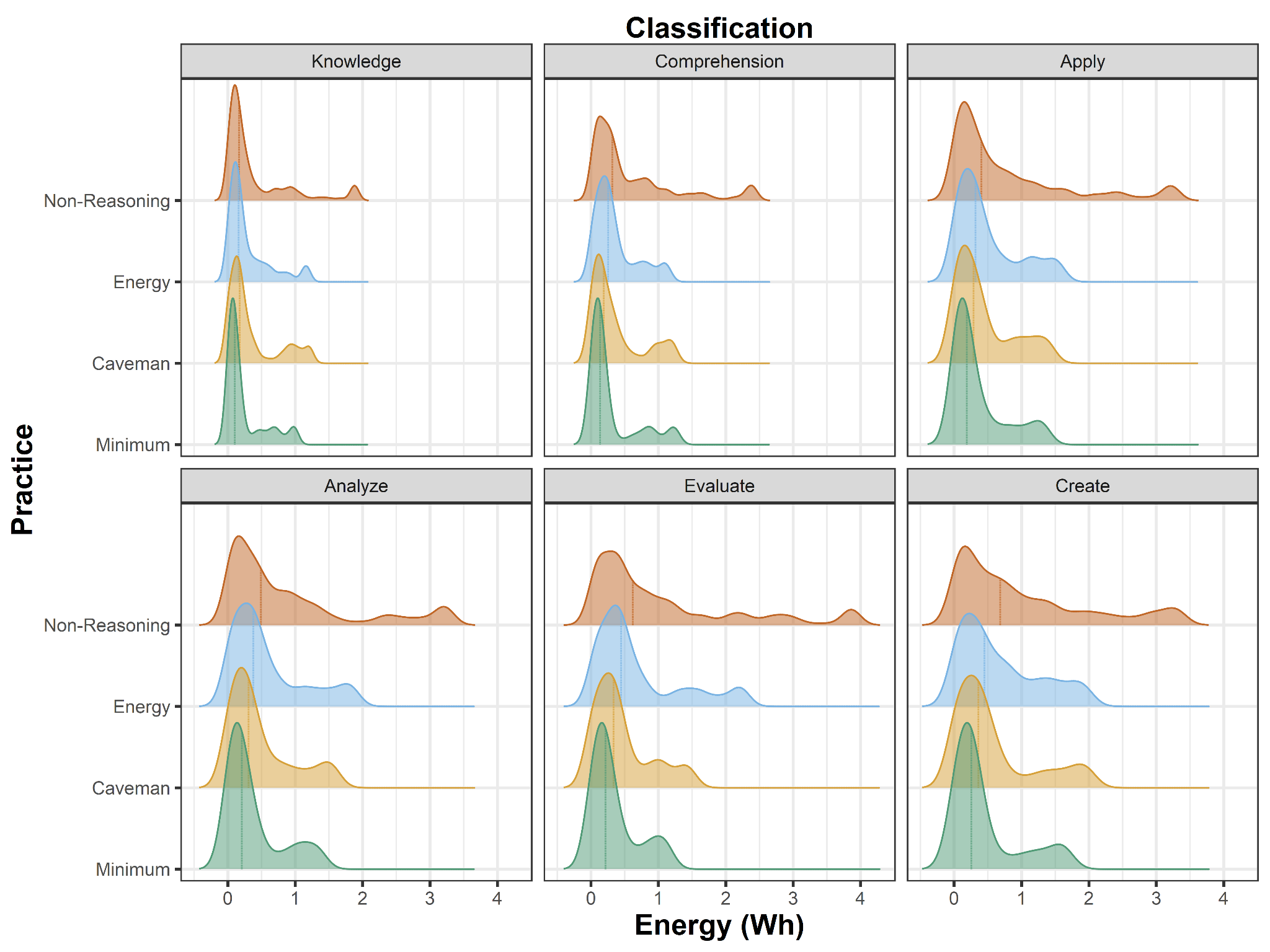}
\caption{Comparison of energy use between baseline and prompting practices.}
\label{fig:4}
\end{figure}

Comparing median baseline energy usage with three prompt-based practices---caveman, energy efficient persona, and minimal output---we find that these practices achieve energy reductions of up to 65\% relative to the baseline, depending on the cognitive complexity of the question and the practice in question (Figure~\ref{fig:4}). We find that the minimum output prompting strategy consistently yields the greatest reduction in energy consumption for all question types (38\% in knowledge type questions to 63\% in creation type questions). Caveman prompting caused the second largest reduction for most task types (40\% for Comprehension type questions to 47\% in Creation type questions) with the exception of Knowledge type questions where we saw a 5\% increase in energy use, on average. Finally, the energy efficient persona generally saw the least energy reduction with 4\% in Knowledge type questions to 35\% in Creation questions.

\begin{figure}[htbp]
\centering
\includegraphics[width=0.85\textwidth]{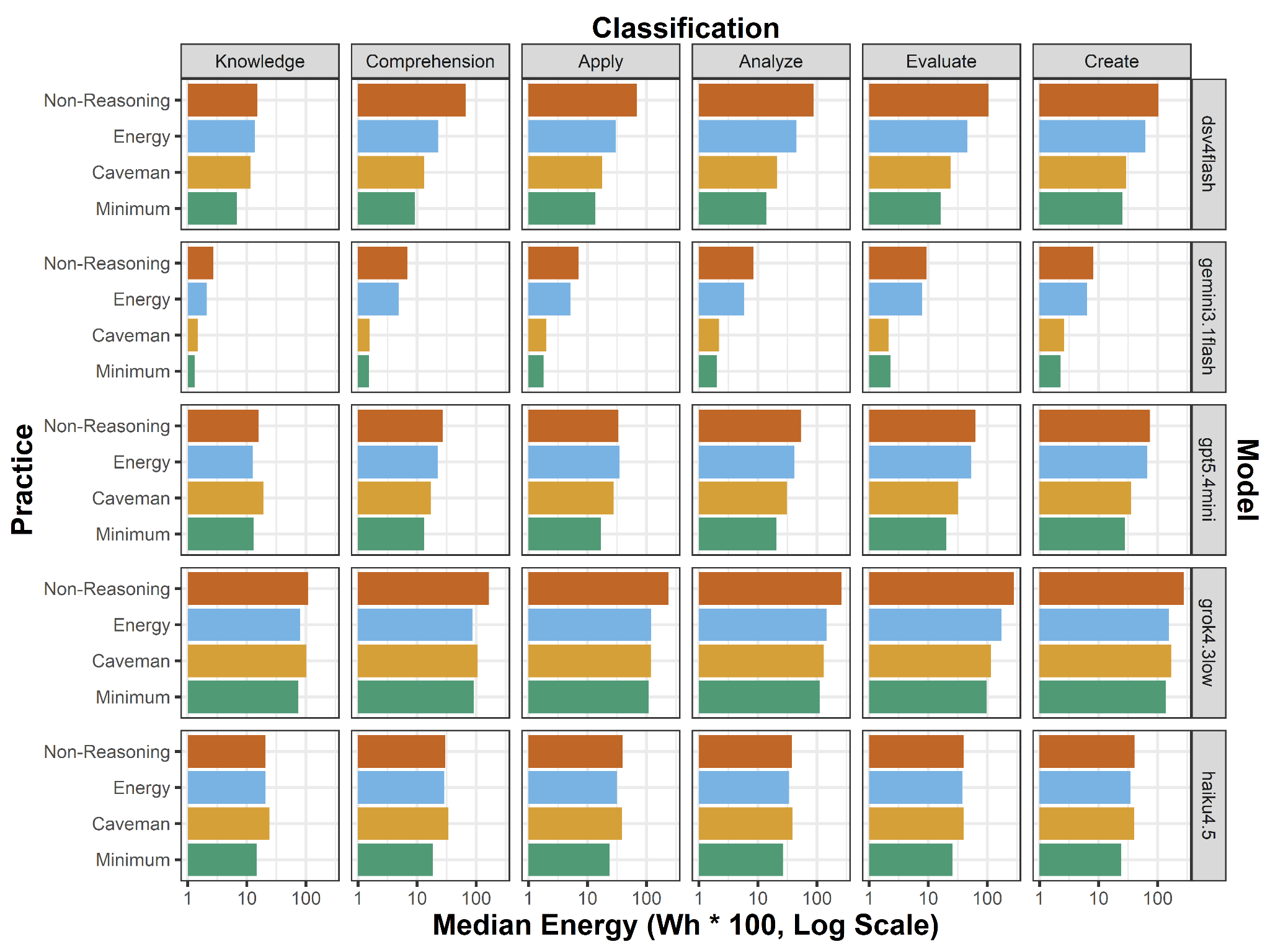}
\caption{Energy consumption for each prompting practice for classification and models.}
\label{fig:5}
\end{figure}

We explored these differences by specific model and found similar results (Figure~\ref{fig:5}), although the effectiveness of prompting techniques on a given task type differed between models. For example, the energy efficient persona was the most effective modification for comprehension type questions on Grok 4.3-low but was the least effective for comprehension type questions on DeepSeek V4 Flash, Gemini 3.1 Flash, and GPT 5.4-mini. Similarly, whereas caveman prompting caused higher energy consumption for most task types on Haiku 4.5, it caused reductions for all task types in every other model except GPT 5.4-mini. Across all task types and prompting techniques, Gemini 3.1 Flash had substantially lower estimated energy consumption than any other model. That different models experienced marked differences in outcomes suggests that the ``best'' practice may change over time, but individual behaviors capable of affecting AI energy consumption exist for all models tested.

\begin{figure}[htbp]
\centering
\includegraphics[width=0.75\textwidth]{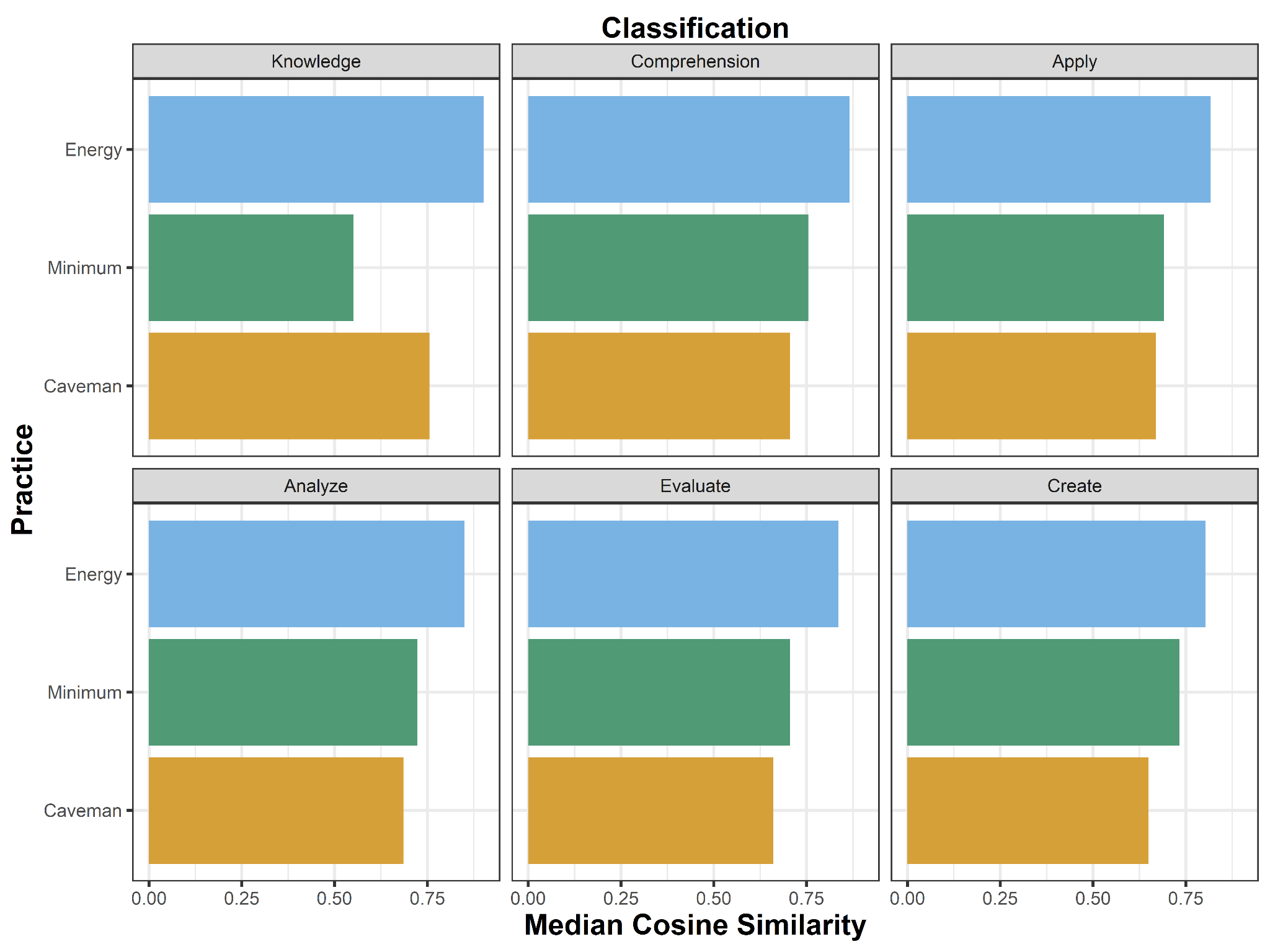}
\caption{Comparison of semantic similarity between baseline and prompting practices.}
\label{fig:6}
\end{figure}

We also analyzed the semantic similarity between responses generated under each prompt-based practice and those generated under the baseline (non-reasoning) models (Figure~\ref{fig:6}). Although the minimum output and caveman prompts produced the largest reductions in energy consumption, they were also associated with lower cosine similarity to the baseline responses (0.64 to 0.75 for caveman and 0.55 to 0.73 for minimum). By contrast the energy efficient persona prompt maintained substantially higher semantic similarity to the baseline, with median cosine similarity between 0.8 and 0.9, while still reducing energy consumption across all task categories. These results suggest that, for a retail user, the energy efficient persona practice may be most effective at preserving semantic similarity to baseline outputs while the minimum answer prompt is generally most effective at reducing energy consumption.

% ============================================================
\section{Discussion}
% ============================================================

The rapid growth of AI-related electricity demand is often framed as a supply-side challenge: how rapidly technology firms and electric utilities can build enough generation, transmission and computing infrastructure to meet rising demand. This study, however, has investigated the other side of the AI-electricity usage equation: demand-side consumption of GenAI tools like LLM-based chatbots that are quickly becoming ubiquitous. The energy demand of commercial LLMs is therefore shaped by millions of small user-side decisions, including the models' users select, whether they utilize reasoning capabilities, the length of outputs, and how they structure prompts. The central question of this study is whether user-side behavioral and prompting modifications can meaningfully reduce the energy consumption of GenAI/LLM chatbot use.

This reframing of the energy impacts of AI usage from the demand and user perspective matters because some of the relevant behavior modifications are simple to understand and immediately accessible to retail users. Across five commercial providers and six categories of task complexity, we find that switching from non-reasoning to reasoning models substantially increases estimated energy consumption, often with limited changes in semantic output. This finding suggests that the routine use of reasoning models for most tasks is difficult to justify from an energy-efficiency or cost perspective. At the same time, the results show that prompt-based practices can reduce energy consumption without requiring technical expertise or changes to model infrastructure. Among the practices tested, the energy-efficient persona prompt provides the highest degree of semantic similarity, and the minimum response provides the greatest average energy reduction.

\subsection{Curving AI's energy consumption}

Prior research has established a close relationship between output length, inference time and energy consumption in LLM use, particularly for open-source models \citep{Li2024,Wang2025,Wilkins2024,Husom2024,Caravaca2025,Podder2026,Adamska2026}. Similar assumptions have been subsequently applied to commercial LLMs through predictive models such as EcoLogits---a suite of open-source tools for estimating the environmental footprint of GenAI models in the inference stage \citep{Rince2025}. In this sense, the energy reductions we observed across prompt-based practices and modifications are consistent with the expectation that shorter responses require fewer output tokens and lower inference time. This learned association between output length and tokens could provide a plausible explanation for why our ``energy-efficient'' persona yielded lower energy consumption compared to the baseline scenario \citep{Husom2024}. Considering the large corpus of work that is used for training all of these models, including scientific repositories like arXiv, it is not surprising to find that the models semantically associate energy-efficiency to brevity, reduced answer length, and changes in the tone of the answer that lead to an overall lower response time and output token count such as the changes we observed with energy-efficient persona prompt.

Although the semantic association may help explain why all three prompt-based practices reduced output length and subsequently token output, the mechanism through which the energy-efficient persona prompt preserved higher semantic similarity remains unclear. One plausible explanation is that the minimal answer and compressed syntax prompts impose stricter constraints on the model's response. By contrast, the energy-efficient persona prompt appears to provide a softer optimization signal to encourage the model to reduce unnecessary verbosity while retaining more of the substantive content of the original answer, as we observed in the Knowledge-type questions, where even relatively simple factual answers may benefit from more contextualization. Further research should examine how different prompt instructions mediate the relationship between output length, semantic similarity, and energy consumption, and whether this relationship can be leveraged to increase energy efficiency in the use of LLMs.

\subsection{The scale of demand-side energy saving in AI use}

Investment in data centers, airport-sized facilities packed with cutting edge computer chips that execute the computationally intensive GenAI programs, is expected to exceed \$725 billion in 2026 and \$7 trillion cumulatively by 2030 \citep{Sullivan2026,Noffsinger2025}. Electricity, and lots of it, is the key input for data centers. Operating at peak capacity, hyperscale data centers demand more power than most cities in the United States and consume more electricity than 800,000 households in a year.\footnote{Most hyperscale data centers operate at the gigawatt scale --- though the largest proposed data centers consume upwards of 9GW.} AI companies are expected to spend more on generation capacity by 2030 than 51 of the largest utilities in the United States combined \citep{Sullivan2026,Powerlines2026}. Reducing the energy demand of AI users can help flatten the electricity curve, thereby limiting the need to invest in new generation capacity. The effect would be significant savings in energy, emissions, and dollars. Though broader systematic changes may have more substantial effects, as mentioned before, the benefits of these practices are tangible and can be seen immediately.

In aggregate, the potential scale of the energy savings of demand-side LLM-prompting modification is substantial. To illustrate, if all the 324.9 million people in the US with access to the internet ask non-reasoning LLMs 20 questions every day of the year, the energy usage is between 7.1--177.7 million kWh---enough to cover the annual household energy demand for up to 16,700 homes in the US. These estimates include only the active inference time required to generate model responses and do not account for associated energy costs, such as idle energy consumption or non-GPU/CPU energy use in data centers. Even with these limitations, our results are broadly consistent with estimates reported in prior work, including \citet{Jegham2025}, Google's self-reported estimate of 0.1 Wh per query for active computing in \citet{Elsworth2025}, and the range of estimates for GPT models reported by \citet{Oviedo2026}.

Using our energy-efficient persona prompt as an example, if the 324.9 million people in the US with access to the Internet included the text ``You are an energy efficient LLM designed to minimize energy consumption from your use without reducing response quality'' in their prompting, it would save enough energy to cover the annual household electricity consumption of up to 7,200 US households with negligible impacts on response quality \citep{EIA2023}.

\subsection{Moving energy-efficient AI forward}

Adopting these best practices aligns with key incentives for all parties involved. Users get more direct, quicker responses without losing quality. They can also help users act on preferences for more sustainable technology use. Employers, enterprises, or other organizations that pay for the AI usage of many individuals could see lower costs due to increased reliance on cheaper non-reasoning models and fewer output tokens (which are generally several times more expensive than input tokens). Model developers can reduce costs for users who use LLMs for free or with a monthly subscription price. Model developers could even add one of these prompt techniques into the system prompt for free users to save significant costs. Data centers can cut electricity costs and reduce wear on expensive GPUs while still meeting the needs of their tenants. Electric utilities could be more capable of meeting their regulatory requirements without spending billions of dollars on new infrastructure, the costs of which will be passed onto households. It could even help meet the federal government's priority to rapidly deploy AI by reducing the need for time consuming and expensive new electricity infrastructure.

Research has demonstrated the lack of information available to users and researchers about the energy use and environmental impacts of AI \citep{Vandenbergh2026}. Although precise estimates of AI energy use and the effects of various practices are difficult to produce, rough approximations are possible. In addition, for the purpose of assessing whether retail behavior change can have important effects on the timing and amount of electricity demand, precise estimates are less important than comparisons among relevant behaviors. In the face of complex systems and limited information, ``muddling through'' is often the most functional approach \citep{Lindblom1959}. The precise energy results for the behaviors assessed by this research may shift over time with changes in the technology, type of energy supply, and other factors, but this research identifies actions that likely use less energy than the alternatives and are likely to do so for an extended period of time. Based on a review of the literature, the actions also are likely to be widely adopted with well-designed interventions. In addition to adoption by household AI users, corporate and government employers, school systems, and other organizations have incentives to reduce the costs arising from AI energy use and can be expected to include energy efficiency information in policies and communications. As AI researchers and users battle an opaque information ecosystem, these best practices can help inspire net improvements in AI use and serve as a foundation for future research.

% ============================================================
\section{Conclusion}
% ============================================================

This article presents some of the first research assessing the effect that casual user practices can have on AI energy consumption. We find that these practices, easily applicable by even the most basic users, can cause a notable reduction in energy consumption without significant reductions in response quality. This result suggests that demand-side initiatives involving changes in users' behaviour can help to curve the increasing demand of energy for AI products without making changes that would negatively affect model quality.

% ============================================================
\section*{Data availability}
% ============================================================
Dataset with all available questions is available at \url{https://doi.org/10.15139/S3/RMU1TU}.

\section*{Acknowledgements}
The authors want to acknowledge the contributions of Alicia Bao, Gamma Posselt, Rojin Sharma, and Harshita Ahuja in data collection, visualization, and styling.

\section*{Funding}
A.H. and D.M. were supported by a grant from the William and Flora Hewlett Foundation awarded to A.H.

\section*{Author contributions}
All authors contributed to writing---review and editing of the manuscript. Conceptualization, methodology, data collection, and analysis were carried out by D.M., E.T., M.S., A.H., and M.V. Software implementation and experiment deployments were conducted by D.M., J.Z., and J.H. The first draft of the manuscript was written by D.M., E.T. and A.H., and all authors provided comments on all versions of the manuscript. The study was supervised by A.H. and M.V. All authors read and approved of the final manuscript.

\section*{Competing interests}
The authors declare that they have no competing interests.

\bibliographystyle{apalike}
\bibliography{references}

% ============================================================
\appendix
\section{Supplementary Material}
% ============================================================

\begin{figure}[htbp]
\centering
\includegraphics[width=0.85\textwidth]{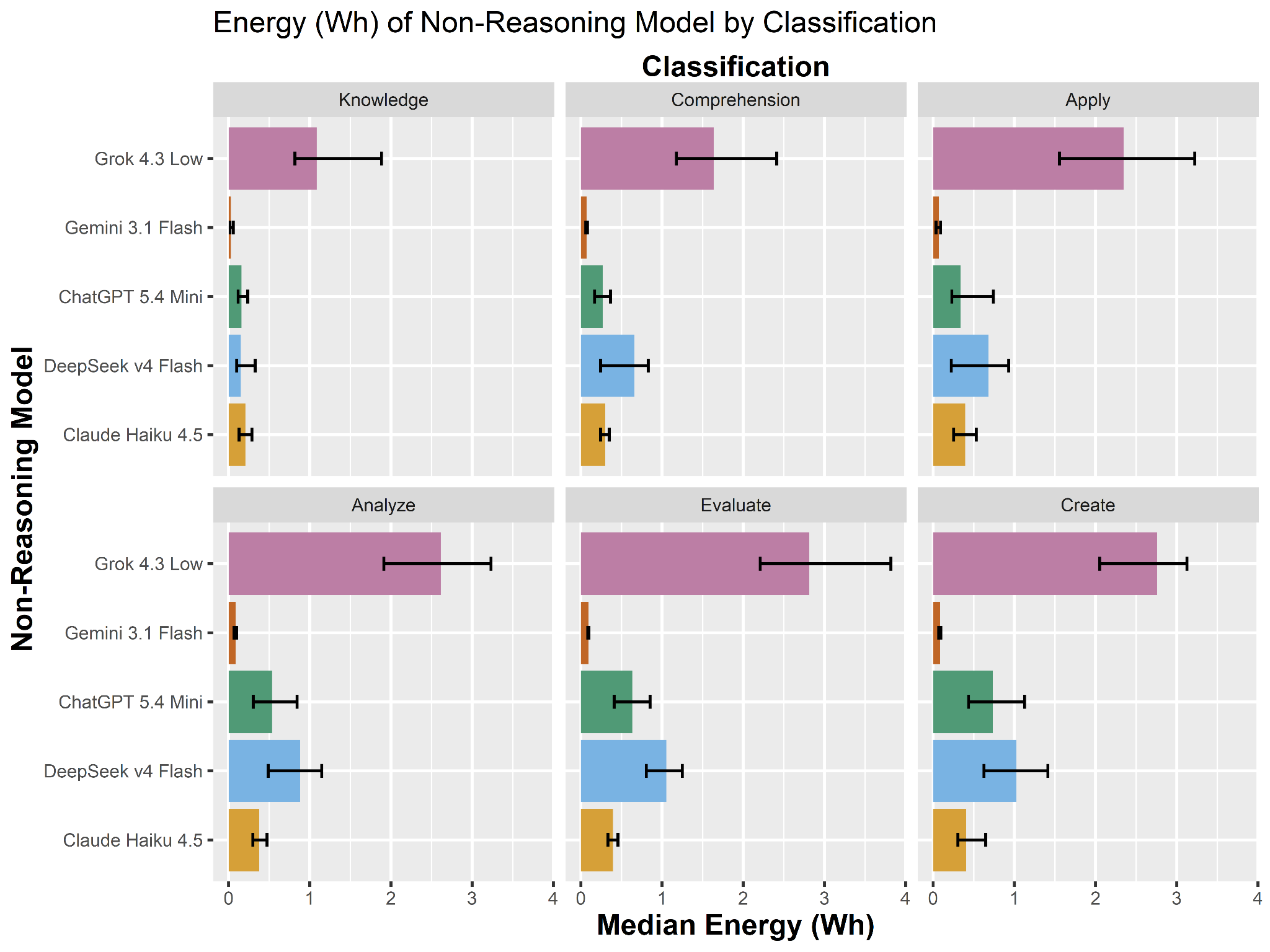}
\caption{Median energy usage (kW) distribution by model and classification type according to Bloom's taxonomy.}
\label{fig:s1}
\end{figure}

\begin{figure}[htbp]
\centering
\includegraphics[width=0.85\textwidth]{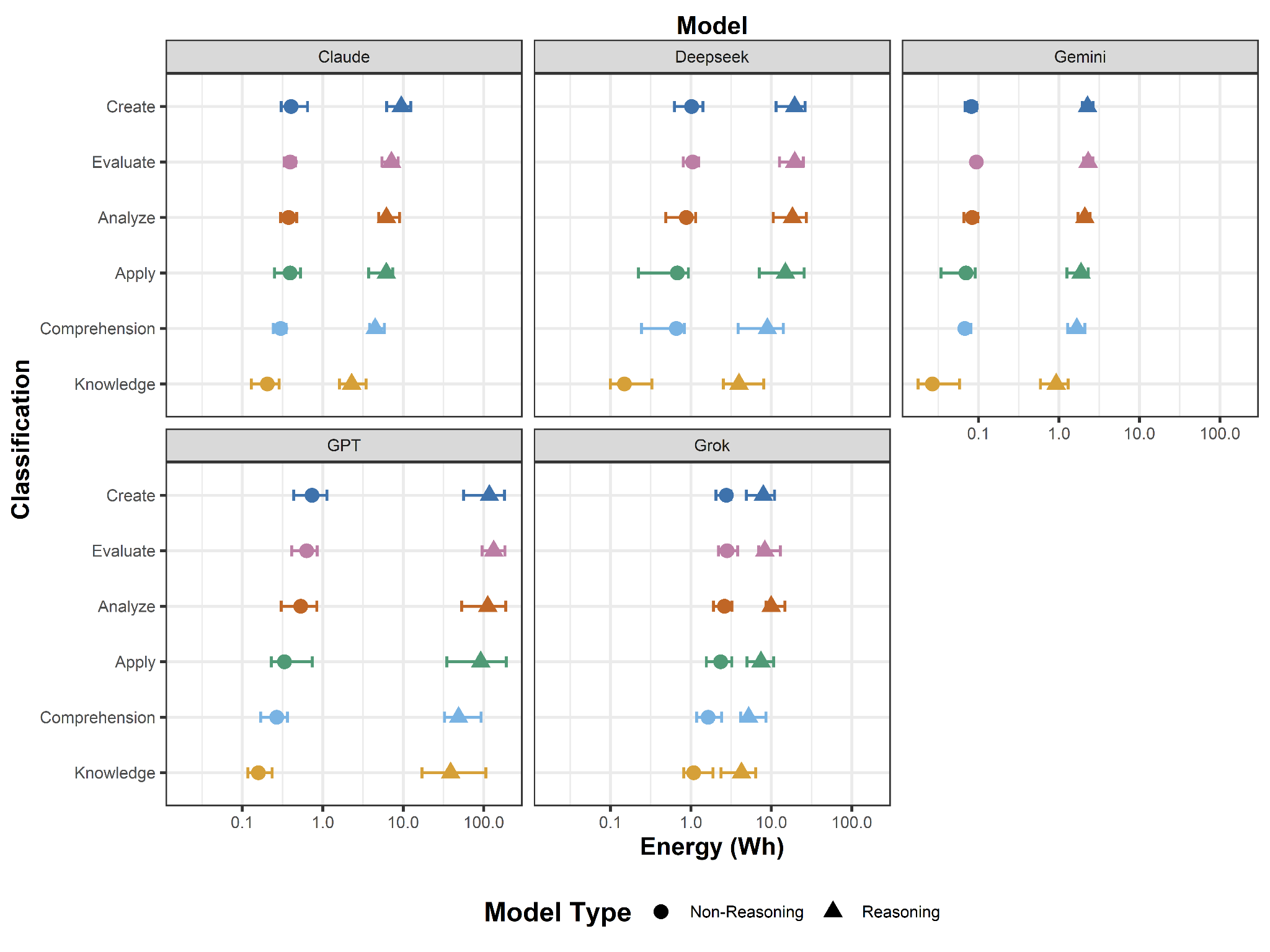}
\caption{Standard deviation of energy use for each model and question classification.}
\label{fig:s2}
\end{figure}

\end{document}